\documentclass[10pt]{revtex4}
\usepackage[dvips,usenames]{color}
\usepackage{latexsym}
\usepackage{amssymb}
\usepackage{amsmath}
\usepackage{epsfig}
\begin{document}

\title{Maxwell demon in Granular gas: a new kind of bifurcation?  The
  hypercritical bifurcation}

\author{M. Leconte, P. Evesque}
\affiliation{Laboratoire MSSMat, \'Ecole Centrale Paris, UMR 8579
  CNRS, Grande voie des vignes, 92295 Chatenay-Malabry, France.}

\begin{abstract}

This paper starts with the investigation of the behaviour of a set of
two subsystems which are able to exchange some internal quantity
according to a given flux function.
%and where an equilibrium is reached when flux equilibrate.
It is found that this sytem exhibit a bifurcation when the flux passes
through a maximum and that its kind (super-critical/sub-critical)
depends on the dissymmetry of the flux function near the maximum. It
is also found a new kind of bifurcation when the flux function is
symmetric: we call it hypercritical bifurcation because it generates
much stronger fluctuations than the super-critical one.
The effect of a white noise
%on this flux
is then investigated.
We show that an experimental set-up, leading to the Maxwell demon in
granular gas, displays all these kinds of bifurcation, just by changing the
parameters of excitation. It means that this system is much less
simple as it was thought.
%This paper investigates the behaviour of two subsystems which are able to
%exchange part of their internal quantity thanks to a flux
%function. This system is described with an analysis of the property of
%the flux function and of the effect of a white noise. A new kind of
%bifurcation, called hyper-critical, is then highlighted where a
%continuous for which the effect of the noise is crucial.
%An experimental attempt to apply our results is also presented.

\end{abstract}

\maketitle

\section{Introduction}

The theory of dynamical systems is a general field of research which
has many applications in different areas, including fluid dynamics
\cite{drazin97}, population dynamics \cite{murray02} or chemical
reaction \cite{mei00} among others.
The idea of this field is to connect physical problems which are
completely different and which pertain in different areas, just
because they are driven by a similar set of equations. Due to this, a
solution obtained in one case can be applied to the other domains,
just by settling the analogy. But to be correct, the method requires
to settle the correct analogy, i.e. the correct change of variables
and boundary conditions, which needs in turn some rigour.
One of the key issue in these problems is
the determination of the stationary states, their stability
(attractors), the dimension of the stable subspace ...
Also one can ask how the solutions evolve when applying different
constrains on the system. This means 
to study the variation of the set of attractors as a
function of external conditions to determine some possible
bifurcation. The way to proceed is well known and it is possible to
characterise the transition between different stationary states thanks
to the bifurcation theory \cite{crawford91,pomeau,manneville}. This
latter allows to describe the dependence of a stationary state with
respect to some experimental parameters for instance.
It has then been pointed out that a single stationary state may
evolve to two or more stationary states as a control parameter
changes. This has led to study the well-known (super-)critical and
sub-critical bifurcations for instance. However, it is also possible
that a stationary state evolves and splits into a continuous set of
stationary 
states but this has not been studied yet. Nevertheless, as we
demonstrate in this paper with a specific and simple example, this
latter kind of
bifurcation. We will even show experimentally a simple example where
all these different types of bifurcations can occur, i.e. super-critical and
sub-critical, plus the hypercritical one. In this case, this last one
will occur when the
system jumps from one kind of bifurcation to the other kind. We will
firstly present the theoretical approach. Then we will present the
classical experiment  of the ``Maxwell demon in granular gas''. And we
will prove that this experiment
can be used to study all these three type of bifurcation. In turn, it
demonstrates that this experiment is much less simple than it was
thought, and that the published explanations are not complete.\\
We will start the paper with the theoretical analysis.
To simplify the concepts, we
consider a system made of two point-like subsystems exchanging a part
of some physical quantity. This quantity is an internal property of
extensive nature. And we investigate its dynamics.
We will focus on a case where the exchange is driven by the
flux functions of each subsystems. Each flux function will be assumed
to depend only on the content of each subsystem (i.e. it does not
depend on the other subsystem).\\
When both subsystems are submitted to the same external conditions,
one expects that both flux functions are identical. A steady state
corresponds to an equilibrium state if it is stable. When the flux
function has a maximum, multiple equilibrium
states can exist. So, a bifurcation occurs when the control parameter
allows the flux to evolve and pass through the maximum.
The kind of bifurcation depends
drastically on the shape of the flux curve. We will see that a new
kind of bifurcation, called hypercritical, can be found
with some special constraints on the
shape of the flux function. It corresponds to the case when a unique
attractor degenerates into a subspace of dimension $d_A$ larger than
$0$. This generates an indeterminacy at some precise value of the
constrains. Apart from this precise constrain, either a single
attractor or a set of two attractors exist. The case which will be
studied here is the case $d_A=1$ but
larger dimension of subset could be envisaged ($d_A=2, 3$) or fractal
subsets too $d_A>0$. Most of the analysis will use a continuum
approach. However, at the end of the theoretical part the effect of
noise will be introduced; this one can be due to the "granular" nature
of the extensive quantity. When this noise is taken into account, the
kind of
bifurcation is particularly ticklish to display and the set of the two
subsystems undergoes a diffusion-like process, which may be solved
using a Langevin's formalism. This diffusion process may broaden the
"point-like" attractor when the attractor is unique, or when the
attraction of the attractors is large compared to the noise agitation
when there are more than one attractor. In this case the steady state
expands over a given volume of the phase space. But the noise can also
generate repeated jumps between the set of attractors when attractors
are discrete in this last case, or even provoke a diffusion among a
continuous or discrete
set of possible attractors. All these configurations will be
considered.\\

In the last part, the paper will revisit the problem of the Maxwell's
demon in granular matter, which can be stated as follows: consider two
halve-containers containing few grains each and connected by a lateral
hole. The system is vibrated and the grains are agitated so that they
can pass from one container to the other one alternately. In practise
one should expect an equi-repartition. However this one is ensured
only when the vibration excitation is large enough, but when the
excitation becomes two small and the dissipation is large enough, one
observes that a container is more filled than the other. This occurs
because the larger dissipation in the more filled container
reduces the
agitation of the grains and the probability of these ones to
escape, while the smaller dissipation in the second container
increases the speed of the grains in this box and favoured the
transfer to the other box. The present experiment will use a single
vibrated container containing some grains, and in which a hole has
been drilled on a side to allow grains to escape. This will allow to
measure the flux function J of grains going out from the box as a
function of the number N of grains in the container. We will see that
the measured flux functions let predict the existence of an
hypercritical bifurcation for some range of parameters. As this
result was not found by previous experiments, numerical
simulations and theoretical approaches, it proves that the problem
was not properly settled and that the correct equations are still not
given. This proves that much work remains to be performed on granular
materials and that much care has to be taken to settle correct
analogies.
%prior to announce a correct understanding of the observed
%phenomena.

The paper is built as follow:
Section \ref{prel} presents the general issue, the equations which
rule the evolution of the
system and the equilibrium condition for which the internal quantity
in each subsystem remains constant.
Section \ref{bifurtheo} presents an analysis of the bifurcation
problem occurring when the internal quantity maximises the flux
function. The hypercritical bifurcation is defined and the condition
of its occurrence characterised. The effect of noise is also
discussed.
Finally, section \ref{exp} is devoted to a convenient experiment to
study the so called ``Maxwell's demon
phenomenon in granular matter'' presented above.
The paper is concluded in section \ref{conclu}.

\section{Preliminaries}\label{prel}

Consider two subsystems, left and right ($l$, $r$), each characterised by
an internal extensive quantity, $x_l$ and $x_r$ and submitted to a set
of external conditions, $y_l^{(i)}$ and $y_r^{(i)}$. We assume first
that the set of the two subsystems is closed such that:
\begin{equation}
  x_r+x_l=x_{tot}=2x_0=cste.\label{contrainte}
\end{equation}
Also the exchange between the two subsystems is ruled thanks to two
flux functions, $J_l$ and $J_r$ associated to each subsystem. These flows
are assumed to depend on the external parameters, $y_l^{(i)}$ or
$y_r^{(i)}$, applied to each container and on its content $x_l$ or
$x_r$ only.
Thus, this paper aims at describing the evolution of such subsystems,
at finding if stationary or equilibrium states exist, if they are
stable (equilibrium states) or not and at studying the possibility of
exhibiting some bifurcations when external parameters, $y_l^{(i)}$ or
$y_r^{(i)}$, evolve.

As an example, one can consider the ``granular Maxwell demon''
\cite{schlichting96,eggers99} where two containers connected by a slit
are vibrated. These containers are partially
filled with macroscopic particles and can exchange some
particles across the slit. Here, the internal quantities are the
number of particles in each box while the external conditions are the
dimension of each box, the position and size of the slit, the
frequencies and amplitudes of vibration for each box, the gravity and
so on. In this peculiar case, it is known that particles equipartition
is observed for intense enough vibration, but that it breaks at small
enough vibration excitation, displaying a threshold and a
bifurcation.\\

Coming back to the more general case, the time evolution of the system
is given by the time evolution of $x_l$ and $x_r$ which writes:
\begin{equation}
  \frac{dx_l}{dt}=J_r(x_r)-J_l(x_l). \label{eq0}
\end{equation}
And equation \eqref{contrainte} imposes:
\begin{equation}
  \frac{dx_r}{dt}=-\frac{dx_l}{dt}. \label{eq0b}
\end{equation}
Both subsystems are stationary (and then in equilibrium but not always
stable) if $\frac{dx_{l,r}}{dt}=0$, which imposes in turn that their
flux are equal:
\begin{equation}
  J_l(x_l)=J_r(x_r).
  \label{equilibre_flux}
\end{equation}
One has now to determine whether this steady/equilibrium state
($x_{l,0}$, $x_{r,0}$) is stable or not. It is stable if any small
perturbation $\delta x$ of $x_{l,0}$ and $x_{r,0}$
(i.e. $x_l=x_{l,0}+\delta x$ and $x_r=x_{r,0}-\delta x$) decreases
spontaneously with time. It is unstable on the contrary.
A first order expansion gives:
\begin{eqnarray}
  \frac{d(x_{l,0}+\delta x)}{dt}&=&J_r(x_{r,0}-\delta x)-J_l(x_{l,0}+\delta x),\\
  \Rightarrow \frac{d\delta x}{dt}&\approx&-\delta x \left [\frac{dJ_r}{dx}(x_{r,0})+\frac{dJ_l}{dx}(x_{l,0})\right ],\\
  \Rightarrow \frac{d\delta x}{dt}&\approx&-\alpha \delta x. \label{equi}
\end{eqnarray}
Where
$\alpha=\frac{dJ_r}{dx}(x_{r,0})+\frac{dJ_l}{dx}(x_{l,0})$.
Consequently, a steady/equilibrium state is stable when $\alpha>0$ and
unstable when $\alpha<0$.\\

For the sake of simplicity, we will limit hereafter the formulation to
two systems with some symmetries, but generalisation is
straightforward. So, we will assume that both systems are identical
(same size, slit at the same position, ...)
and are submitted to equivalent external conditions. It means in
particular that $J_l$ and $J_r$ depend only on $x_l$ and $x_r$
respectively and on a unique set of external conditions. This imposes
also that $J_l(x)=J_r(x)=J(x)$.\\
One would like now to know if a stable equilibrium state can become
unstable when varying a control parameter. This leads to determine
under which condition a bifurcation could occur and which kind of
bifurcation could be obtained from this formalism.

%\section{Theoretical part}\label{part1}

\section{Bifurcation Analysis}\label{bifurtheo}

\subsection{Classical analysis}\label{classique}

Let us assume that the two subsystems contain both $x_0$
initially. From equation \eqref{eq0b} and according to equation
\eqref{equilibre_flux}, as soon as the two subsystems are identical and
are submitted to the same constrains, the state defined by
$x_l=x_r=x_0$ is a steady state. According to equation \eqref{equi}, this
state remains stable as long as the first derivative
of $J$ at $x_{l/r}$ is positive  that is, as long as $J$ increases
with $x$ at $x_0$. This solution becomes unstable as soon as $J$ starts
decreasing. In this case, a new solution has to be found which
satisfies $x_l+x_r=2x_0=cste$ and $J(x_l)-J(x_r)=0$. This is the jump
from the unique solution to the set of two different solutions which
will interest us during this section. It occurs when $dJ/dx$ passes
through $0$.\\

Generally, as $J$ is an outgoing
flux, one expects that $J$ increases at small $x$ starting from
$J(0)=0$. So, the jump/bifurcation occurs when $J$ passes through a
maximum. However, it may occurs that $J$ starts from a non null value
and passes through a minimum in some cases but the analysis is similar
to the one made in the following when $J$ increases at small $x$.\\
Note also that we will consider in this paper only the flux function for
which the first derivative could cancel at most one time and the case
where $J$ is infinitely derivable.\\

So, turning back to the most probable case for which
$J$ increases at small enough $x$, i.e. $\frac{dJ}{dx}>0$ at small $x$,
it means that a transition occurs when the flux function reaches its
maximum value $J(x_m)$.
This leads to study and determine the kind of bifurcation which occurs
around $x_{l,r}=x_m$. The parameter $v$ which controls the
distance to the threshold is $v=x_l+x_r-2x_m$ and can be used to
study the bifurcation nature. In this case, the preserved quantity
$x_l+x_r$ writes $x_l+x_r=2x_m+v$. It is also convenient to
introduce the asymmetry parameter $u=x_l-x_r$ as the order
parameter. So writing $J$ as a function of the distance to the maximum
and using a Taylor expansion, one obtains:
\begin{eqnarray}
  J(x_l-x_m)&=&J(x_m)+\sum_{k=1}^{\infty}\frac{(x_l-x_m)^k}{k!}
  \frac{d^kJ}{dx^k}(x_m),\nonumber \\
  J(x_r-x_m)&=&J(x_m)+\sum_{k=1}^{\infty}\frac{(x_r-x_m)^k}{k!}
  \frac{d^kJ}{dx^k}(x_m).
  \label{taylor_max}
\end{eqnarray}
We use the notation hereafter
$J^{(k)}(x_m)=\frac{d^kJ}{dx^k}(x_m$.
By hypothesis, $J^{(2)}(x_m)=\frac{d^2J}{dx^2}(x_m)<0$ and
$J^{(1)}(x_m)=\frac{dJ}{dx}(x_m)=0$ is the bifurcation
threshold.

So, using the change of variable:
\begin{eqnarray}
  \frac{u+v}{2}&=&x_l-x_m, \label{chgt1}\\
  \frac{v-u}{2}&=&x_r-x_m. \label{chgt2}
\end{eqnarray}
The dynamics equations becomes at third order:
\begin{eqnarray}
  \frac{du}{dt}&=&-u\left (
  vJ^{(2)}(x_m)+\frac{v^2}{4}J^{(3)}(x_m)+...\right ) \nonumber \\
  &&-u^3\left (\frac{1}{12}J^{(3)}(x_m)+...\right ) + ... \nonumber\\
  \frac{dv}{dt}&=&0.
  \label{evol_temps}
\end{eqnarray}
%Where all the derivative are computed at $x=x_m$,
$J^{(1)}(x_m)=0$ and $J^{(2)}(x_m)<0$ by hypothesis.\\
The solution of equations \eqref{evol_temps} at first order is:
\begin{equation}
  u(t)\sim \exp{-(vJ^{(2)}(x_m)t)}. \label{solu}
\end{equation}
As $J^{(2)}(x_m)<0$, the solution defined by equation \eqref{solu} is
stable for $v<0$ and unstable for $v>0$
and one deduces the typical time $\tau$ of evolution
$\tau=\left (vJ^{(2)}(x_m)\right )^{-1}$. It is worth noting that $\tau$
tends to infinity as $v$ tends to $0$. This is the so-called
``critical slowing down'' and it means that the equilibrium needs an
infinite time to be reached. To get the correct time behaviour at
$v=0$, one shall expand equation \eqref{evol_temps} to higher order in
$u$. This leads to a power law relaxation instead of an exponential
one.\\

A stationary state corresponds to $\frac{du}{dt}=0$. Thus, a solution
to equation \eqref{evol_temps} with $\frac{du}{dt}=0$ is:
\begin{eqnarray}
  u&=&0. \label{unul}
\end{eqnarray}
As already told, this solution is stable if $v<0$ and unstable if
$v>0$. It means that $x_l=x_r$ is a stable equilibrium state as long
as $x_l+x_r<2x_m$.
So, when $v>0$ the solution $u=0$ is unstable. To get the behaviour,
one has to take account for higher order term in the
development.\\

\subsubsection{Case $J^{(1)}(x_m)=0$, $J^{(2)}(x_m)<0$, $J^{(3)}(x_m)>0$}

\begin{figure}[!h]
  \begin{center}
    \begin{minipage}{65mm}
      \psfig{file=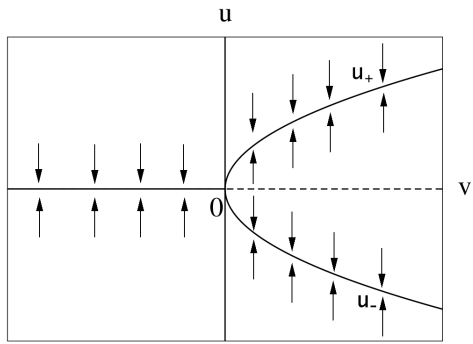,width=65mm,angle=0}
    \end{minipage}
    \caption{\small Solutions of equations \eqref{evol_temps}
    for $J^{(3)}(x_m)>0$. Full lines show
    the stable solution
    while dashed line shows the unstable ones. Arrows describe the
    time evolution of a perturbation $\delta u$. Super-critical bifurcation
    when $v$ becomes positive.}
    \label{bifur_crit}
  \end{center}
\end{figure}

Limiting to the third order, this gives two new
solutions for equation \eqref{evol_temps} when $J^{(3)}(x_m)>0$
and $v>0$:
\begin{eqnarray}
  u_{\pm}&=&\pm \left (-v
  \frac{J^{(2)}(x_m)+... }
       {\frac{1}{12}J^{(3)}(x_m)+...} \right )^{1/2},\nonumber \\
  &=&\pm\sqrt{\mu}. \label{unonul}
\end{eqnarray}
One can show that these solutions are stable. As the two new solutions
do not exist at $v<0$, the bifurcation is the one described in figure
\ref{bifur_crit}, with a parabolic branching at $v>0$. This is
characteristic of a super-critical fork bifurcation (because the new
solutions are stable). In the present case, it occurs at $J^{(1)}(x_m)=0$, with
$J^{(2)}(x_m)<0$ and $J^{(3)}(x_m)>0$.

The two solutions appear also when $J^{(3)}(x_m)>0$ and $v<0$ but
this case will be studied later. Let us simply mentioned that they are
unstable.

\begin{figure}[!h]
  \begin{center}
    \begin{minipage}{65mm}
      \psfig{file=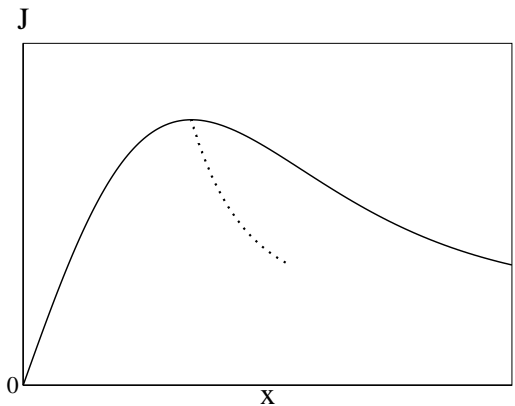,width=65mm,angle=-90}
    \end{minipage}
    \caption{\small An example of the shape for $J$ corresponding to a
    super-critical bifurcation (i.e. with $J^{(3)}(x_m)>0$). The right wing
    shall be broader than the left wing. The dashed line is the
    location of the middle of horizontal secant.}
    \label{critbif}
  \end{center}
\end{figure}

To conclude with the case
$\{J^{(1)}(x_m)=0,J^{(2)}(x_m)<0,J^{(3)}(x_m)>0\}$, when $v$ increases
from $v<0$ to $v>0$, the solution $x_l=x_r$ then breaks following a
super-critical bifurcation with two symmetric and stable solutions $u_\pm$
shown on Figure \ref{bifur_crit}. Figure \ref{critbif} shows a typical
shape of $J$ when $J^{(2)}(x_m)<0$ and $J^{(3)}(x_m)>0$. It is
characterised by a left wing steeper than the right wing. This allows
the $x$ position of the middle of an horizontal secant at given $J$ to
increase as $J$ decreases, which is the requirement to get the
super-critical bifurcation. Indeed, in the case $2x>2x_m$, any steady
solution above $x_m$ requires to find two solutions $x_1$ and $x_2$,
such as $J(x_1)=J(x_2)$ and $x_1+x_2=2x>2x_m$. Such a possibility
exists only if the middle of the horizontal secant at a given $J$
increases when $J$ decreases. This is also what $J^{(3)}(x_m)>0$ means
when $J^{(1)}(x_m)=0$ and $J^{(2)}(x_m)<0$.\\

\subsubsection{Case $J^{(1)}(x_m)=0$, $J^{(2)}(x_m)<0$, $J^{(3)}(x_m)<0$}

\begin{figure}[!h]
  \begin{center}
    \begin{minipage}{65mm}
      \psfig{file=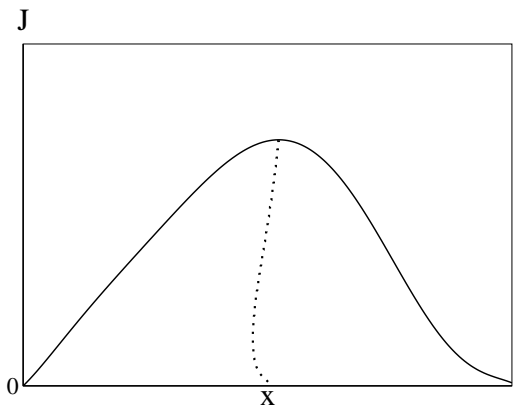,width=65mm,angle=-90}
    \end{minipage}
    \caption{\small Typical shape of $J$ leading to a sub-critical
    bifurcation (with $J^{(3)}(x_m)<0$). The dashed line corresponds to
    the location of the middle of horizontal secant.}
    \label{sscritbif}
  \end{center}
\end{figure}

When $J^{(3)}(x_m)<0$, there is no solution to equation
\eqref{evol_temps} with $\frac{du}{dt}=0$ for $v>0$. This is due to the
fact that the position $x_{mid}$ of the middle of the secant at a
given $J$ increases with $J$ when the right wing of the $J$ curve is
steeper than the left one, as it is exemplified in Figure
\ref{sscritbif}. So, to get a
solution one shall expand equation \eqref{taylor_max} at higher order in
the vicinity of $v=0$ in equation \eqref{eq0}. Furthermore, owing
to the fact that the dynamics of the system is controlled by the
subtraction of two flows, the $u^4$ term cancels (as the $u^2$ term
had disappeared in equation \eqref{evol_temps}) and one shall develop to
fifth order.

Using the change of variable (\eqref{chgt1} and \eqref{chgt2}) and
applying these expansions lead to the following dynamics equation:
\begin{eqnarray}
  \frac{du}{dt}&=&-u\left (
  vJ^{(2)}(x_m)+\frac{v^2}{4}J^{(3)}(x_m)+\frac{v^3}{24}
  J^{(4)}(x_m)+\frac{v^4}{192}J^{(5)}(x_m)+...\right ) \nonumber \\
  &&-u^3\left
  (\frac{1}{12}J^{(3)}(x_m)+
  \frac{v}{24}J^{(4)}(x_m)+\frac{v^2}{96}J^{(5)}(x_m)+...\right
  ) \nonumber \\
  &&-u^5\left
  (\frac{1}{960}J^{(5)}(x_m)+... \right )+... \label{evol_temps2}
\end{eqnarray}

A possible stationary state ($\frac{du}{dt}=0$) of equation
\eqref{evol_temps2} is $u=0$ but it is only stable if $v<0$ in the
vicinity of $v=0$. It is unstable for $v>0$. The other possible
solutions are symmetric compared to $u=0$. Furthermore, equation
\eqref{evol_temps2} with $\frac{du}{dt}=0$ accepts three solutions
when $v>0$ (two stable and one unstable), five solutions in the range
$v\in [v_{min}<0;0]$ (three stable and two unstable) and one stable
solution when $v<v_{min}$.

\begin{figure}[!h]
  \begin{center}
    \begin{minipage}{65mm}
      \psfig{file=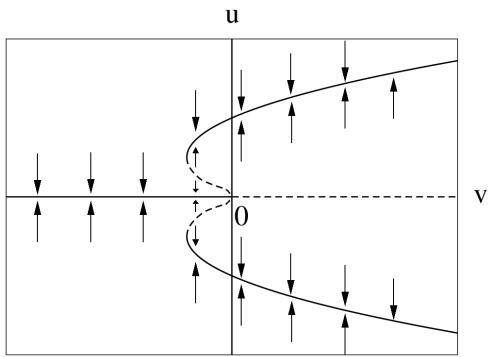,width=65mm,angle=0}
    \end{minipage}
    \caption{\small Solutions of equations \eqref{evol_temps}
    for $J^{(3)}(x_m)<0$. Full lines show
    the stable solutions while dashed lines show the unstable
    ones. Arrows indicate the spontaneous direction of motion of a
    perturbation $\delta u$. This is typical of a sub-critical
    bifurcation.}
    \label{bifur_ss_crit}
  \end{center}
\end{figure}

A typical example of such a case is displayed on Figure
\ref{bifur_ss_crit}. An example of curve $J$ corresponding to this
bifurcation is displayed on Figure \ref{sscritbif}. It is
characterised by a right wing steeper than the left wing, as explained
previously, and by the position $x$ of the middle of a horizontal
secant that increases when $J$ increases beyond a value
depending on its shape.

So, when $v<v_{min}$, $u=0$ is the only (stable)
solution. Then, when $v\in [v_{min}<0;0]$, the solution 
$u=0$ remains stable but there are also two other stable solutions,
both are separated from the $u=0$ solution by an unstable
solution. This is typical of
multistable state, which is known to lead to hysteresis. When $v>0$,
$u=0$ is now an unstable solution and the two symmetric solutions
$u_\pm$ are the two stable
solutions. Hysteresis occurs as follows: we start with $u=0$ at
$v<v_{min}$. So, increasing slowly $v$
from $v<v_{min}$, let the solution $u=0$ unchanged, then increasing
$v$ above $v=0$ forces the jump of $u=0$ to $u_+$ or $u_-$ when $v$
becomes positive. Now, decreasing slowly $v$ from above $0$ to below
$v<v_{min}$ the solution  $u_+$ or $u_-$ evolves slowly till $v$
reaches $v<v_{min}$ where $u$ jumps to $u=0$. This hysteretic
behaviour is typical of a sub-critical bifurcation.\\


For $v=0$, that is $x_l+x_r=2x_m=x_{tot}$, the dynamics equation
writes:
\begin{equation}
  \frac{du}{dt}=-\frac{u^3}{12}J^{(3)}(x_m)
  -\frac{u^5}{960}J^{(5)}(x_m)+... \label{evol_temps_special}
\end{equation}
The dynamics is controlled by $J^{(3)}(x_m)$ and
$J^{(5)}(x_m)$. Return to equilibrium is no more an exponential
decrease but a power law. The
steady solutions of equation \eqref{evol_temps_special} writes at first order:
\begin{eqnarray}
  u&=&0,\\
  u_\pm&=&\pm \left [-80J^{(3)}(x_m)\left (J^{(5)}(x_m)\right
  )^{-1}\right ]^{1/2} \\
  &&\mbox{ if,  }\ J^{(3)}(x_m)J^{(5)}(x_m)<0.\nonumber
\end{eqnarray}
Their stability depends on the
third and fifth derivative of $J$ at $x_m$.
Finally, the comparison of Figures \ref{critbif} and \ref{sscritbif}
shows that the symmetry of $J$ plays an important role on the
bifurcation behaviour.

\subsubsection{Symmetric flux function: $J(x_m-x)=J(x_m+x)$}

A special attention has then to be paid to the
case of a symmetric $J$, i.e. with two symmetric wings, which implies
$J^{(k)}(x_m)=0$ for all odd $k$.
The dynamics equation writes at fourth order in such a case:
\begin{eqnarray}
  \frac{du}{dt}&=&-uv\left
  (J^{(2)}(x_m)+\frac{v^2}{24}J^{(4)}(x_m)\right )\nonumber \\
  &&-u^3\frac{v}{24}J^{(4)}(x_m)+... \label{evol_temps3}
\end{eqnarray}
Where $J^{(2)}(x_m)<0$.\\
Let us first neglect the higher order terms $J^{(4)}(x_m)$ ..., and
consider the problem near $v=0$. As $J^{(2)}(x_m)<0$, the solution
$u=0$ is stable for $v<0$ and unstable for $v>0$.

So, at $v=0$, there is a continuous set of equilibrium states, since
$\frac{du}{dt}=0$ whatever $u$. So, all initial
states are equilibrium states: keeping $v=0$, and
starting from any
definite state $u=u_0$, $u_0$ is a steady state. Suppose now that the
system passes spontaneously to $u_1$ due to some perturbation, in this
case the new $u_1$ state also is the new steady state and the system
will be fully driven by perturbation or uncontrolled
noise that generates transition from $u_0$ to $u_1$.

\begin{figure}[!h]
  \begin{center}
    \begin{minipage}{65mm}
      \psfig{file=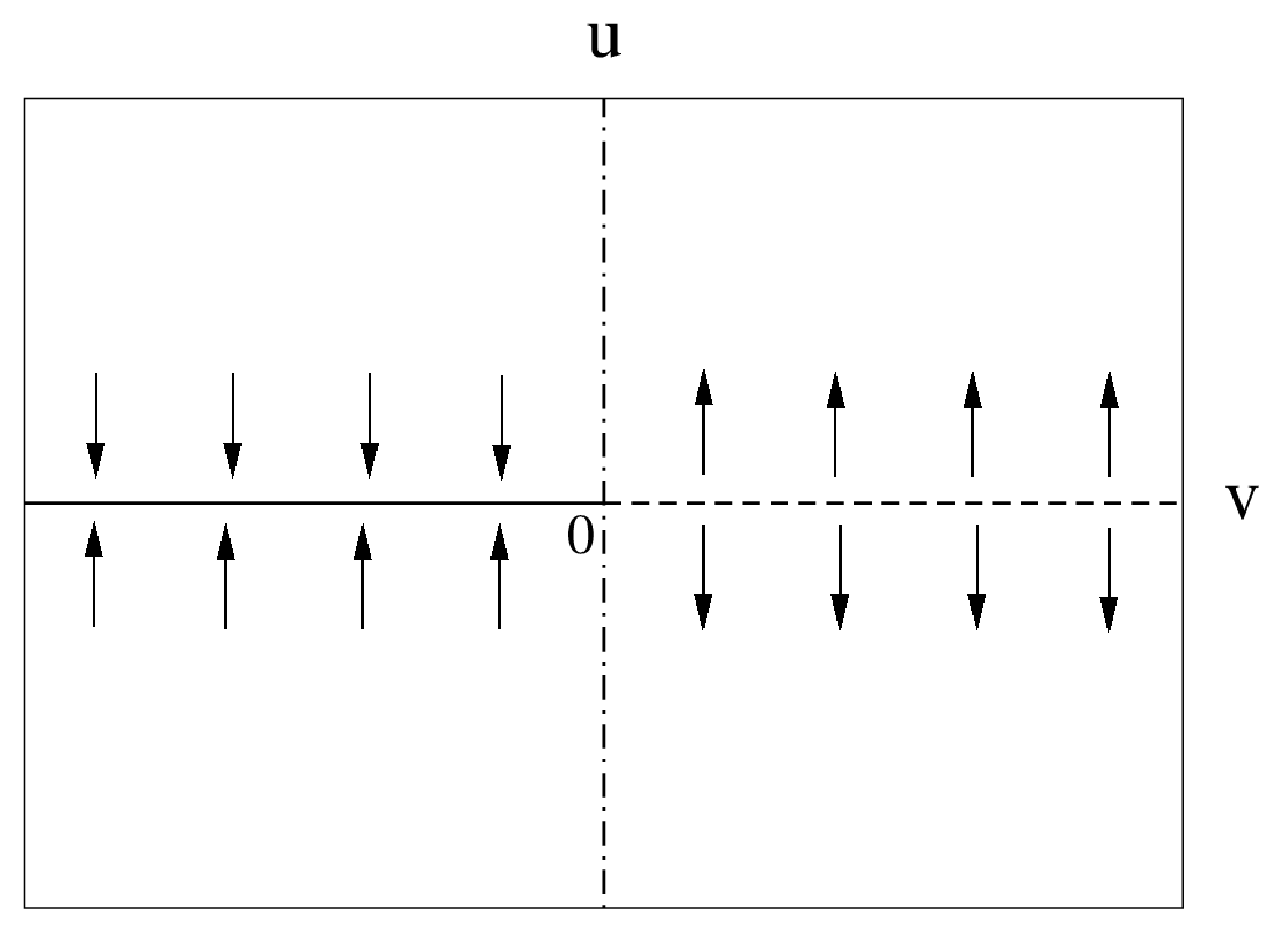,width=65mm,angle=0}
    \end{minipage}
    \caption{\small Solutions of equations \eqref{evol_temps}
    for $v=0$ and $du/dt=0$ for all $u$. Full line shows the stable
    solution while dashed line shows the unstable ones. The dot-dashed
    line shows the set of solutions which are neither stable nor
    unstable. Arrows indicate the spontaneous direction of motion of a
    perturbation $\delta u$. Hypercritical bifurcation.}
    \label{bifur_hyp_crit}
  \end{center}
\end{figure}

\begin{figure}[!h]
  \begin{center}
    \begin{minipage}{65mm}
      \psfig{file=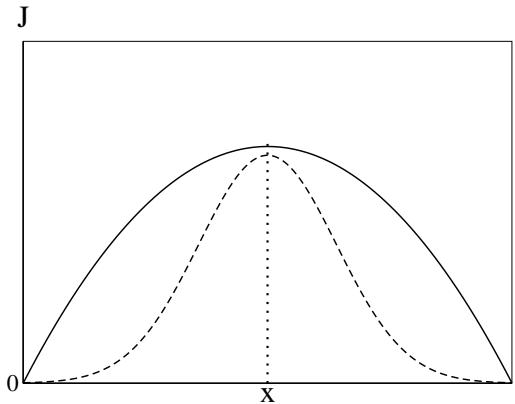,width=65mm,angle=-90}
    \end{minipage}
    \caption{\small Two examples of the shape of $J$ corresponding to a
    hypercritical bifurcation when $v=0$. The curves are symmetric with
    respect to the maximum of $J$, $J(x_m)$. The dashed line is the
    location of the middle of horizontal secant.}
    \label{hypbif}
  \end{center}
\end{figure}

As we will see, such a bifurcation trend is important because it leads
to generate extremely large fluctuation, larger than a critical
bifurcation. We can then call this bifurcation an hypercritical
bifurcation. Its diagram is displayed on Figure
\ref{bifur_hyp_crit}. Figure \ref{hypbif} shows typical functions $J$
which give rise to a hypercritical bifurcation, with the vertical
dotted line representing the location of the middles of horizontal
secant.

This bifurcation is
particularly interesting from an experimental point of view because
its dynamics is completely driven by the existing noise, and because
any extensive quantity is subject to noise due to the
discreetness of nature at some microscopic finite scale.
Indeed, no matter some mechanism exists to absorb
any perturbation or to make the perturbation grow, the
system is then entirely driven by the noise when it is at this working
point even if it covers only equilibrium states.

Does it means that the dynamics of the system at $v=0$ is a simple
diffusion on the $u$ axis? Not completely, because the fluctuations that
generate the motion in the $u$ direction may have specific characteristics
that make the system more complex. For instance, even in the case of a
random process imposed by discrete nature of the flux, as the noise is
often linked to the square root of the magnitude of the flux, one shall
expect already that the diffusion coefficient shall depend on the value of
$u$, because the value of $J$ depends on $u$ (see  Figure
\ref{hypbif}).\\

Since $J(0)=0$, $J(2x_m)=0$ and the flux becomes negative
when $v>0$. This is why figure \ref{bifur_hyp_crit} does not exhibit
solution for $v>0$. However such solutions may appear in this
hyper-critical bifurcation if some dissymmetry of $J$ is introduced at
large $v$ only, i.e. above $x=2x_m-a$ with $a$ being small compared to
$x_m$, so that one will find a solution for $v>0$. This requires of
course that $J(x)$ remains positive for all $x>2x_m-a$ and tends to
$0$ as $x$ tends to infinity.

\subsubsection{Slightly dissymmetric flux function}

If the flux function is slightly dissymmetric in the range $[0;2x_m]$,
that is, if some of the odd derivatives at $x=x_m$ are not null
beyond order $2k+1>5$, the bifurcation is not hypercritical
{\it stricto sensu}. It means that a solution may be given. However
this would require a large/infinite precision especially in the
vicinity of $v=0$ to maintain its
state at the predicted value. On the contrary, let us assume some
existing intrinsic fluctuations. It means that the system may explore
different states with time, and the wider the range explored the
nearer from the symmetry the $J$ function. This is because the
dissymmetry of the $J$ curve is
linked to the derivatives of small order $(2k+1)$ at the
maximum. So, one may expect that in the vicinity of $v=0$, the larger
the order $k$ for which $J^{(2k+1)}(x_m)$ is non null, and/or the
smaller the $J^{(2k+1)}(x_m)$ the wider the range of
$u$ over which the system may spread due to fluctuations. It is then
quite important to include the effect of fluctuations in the dynamics
to treat correctly the problem. This will be done in the next sub-section.

\subsection{Effect of the noise}\label{noise}

The previous section was devoted to the bifurcation analysis when
symmetry breaks. Expected behaviour has been described without taking
into account any
noise and it has been found that the kind of bifurcation
depends strongly on some detail of the shape of the flux
function. This makes the solution sensitive to tiny details. On the
other hand, it is experimentally
impossible to completely cancel the noise. So, one has then to
introduce it to describe properly the physics in the vicinity of the
bifurcation.\\
Among the questions to answer, one would like to know if
the system is
able to jump spontaneously from a stable solution to an other
one just because of the existing noise.\\
%{\bf CORRECTION :\\
%(when the stable solutions are discrete.) \`A enlever.\\
%{\bf FIN CORRECTION}\\

This depends probably on
%{(\bf the stability of the solutions and :
%inutile, on dit plus haut que c'est entre solutions stables.)}
the distance between the stable solutions and on the intensity of the
noise. One can also ask how the
system explores the space of stable configurations in the case of a
continuous set of stable states, or how far from a stable state
the system can be when it is driven by a noise of given amplitude. These
questions can be addressed thanks to
the computation of the fluctuation of $x$ around a stable
solution. But this requires also to model the action of the noise.

Let us start with the last problem and consider a stable state $(x_l;x_r)$
and some perturbation $\delta x$ , i.e. $(x_l+\delta x,x_r-\delta
x)$. Without perturbation, its dynamics writes:
\begin{equation}
  \frac{dx_l}{dt}=J(x_r)-J(x_l)=0.
\end{equation}
Let us slightly perturb this equation and use a Taylor expansion at
first order:
\begin{eqnarray}
  \frac{d(x_l+\delta x)}{dt}&=&J(x_r-\delta x)-J(x_l+\delta x),\\
  \Leftrightarrow \frac{d\delta x}{dt}&\approx&-\delta x
  \left ( J^{(1)}(x_l)+J^{(1)}(x_r)\right ),
  \nonumber\\
  \Leftrightarrow \frac{d\delta x}{dt}&\approx&-\alpha \delta x.
\end{eqnarray}
where we have used $\alpha =J^{(1)}(x_l)+J^{(1)}(x_r)>0$ (because it
is a stable state) and $J(x_l)=J(x_r)$.

Let us submit this system to some random white noise of amplitude
$\varepsilon(t)$, of zero mean and of variance $\sigma_\varepsilon^2$,
the system obeys to a Langevin-like equation \cite{reif65}:
\begin{eqnarray}
   \frac{d\delta x}{dt}&\approx& -\alpha \delta x+\varepsilon(t).\label{eqlang}
\end{eqnarray}
A characteristic time is the time $\tau_c$ for the system to exchange
a ``particle'' (or quantum) of the extensive quantity $x$:
$\tau_c=\frac{1}{J(x_l)+J(x_r)}$.
%This time $\tau_c$ is linked to $\tau_c=1/J$ when $J$ is expressed in
%terms of quanta of the extensive quantity.
Assuming that $t$ is always much larger than the time for $x_l$ and
$x_r$ to exchange one particle, $t\gg \tau_c=\frac{1}{J(x_l)+J(x_r)}$,
equation \eqref{eqlang}
can be solved following the classic procedure as in the case of a
Langevin equation and find the variance $\sigma_{\delta x}^2$ of $\delta x$,
as a function of the variance of the noise, $\sigma_{\varepsilon}^2$
\cite{reif65}. We start from the general form of the solution of
equation \eqref{eqlang}:
\begin{equation}
  \delta x(t)=\delta x(0)e^{-\alpha t}+\int_0^t e^{-\alpha
  (t-t')}\varepsilon(t')dt'.\label{solgene}
\end{equation}
As the noise $\varepsilon$ is a zero-mean random variable, $\langle \varepsilon
(t)\rangle =0$, the expectation of $\delta x$ writes:
\begin{equation}
  \langle \delta x(t)\rangle = \delta x(0)e^{-\alpha t}.
\end{equation}
So, the perturbation damps, in average, from any (small) initial value with a
typical time $\tau$:
\begin{equation}
  \tau=\frac{1}{\alpha}.
\end{equation}
Considering now the variance of $\delta x$ around a stable position,
it is defined as:
\begin{equation}
  \sigma_{\delta x}^2=\langle \delta x(t)^2\rangle -\langle \delta
  x(t)\rangle^2.
\end{equation}
Using the square of equation \eqref{solgene}, one obtains:
\begin{equation}
  \sigma_{\delta x}^2=\int_0^{t'}dt'\int_0^t\langle
  \varepsilon(t')\varepsilon(t'')\rangle
  e^{-\alpha(t-t')}e^{-(t'-t'')}dt''.
\end{equation}
As $\varepsilon$ is a white noise, one finds:
\begin{equation}
  \sigma_{\delta
  x}^2=\frac{\sigma_{\varepsilon}^2}{2\alpha}(1-\exp{(-2\alpha t)}).
  \label{langevin}
\end{equation}
So, when $t=0$, the variance is null because $\delta
x(0)$ is known exactly. As $t$ increases from $0$, $\sigma_{\delta x}$
starts to grow as the square root of time under the effect of the
noise. At larger time, the standart deviation saturates and reaches
$\sigma_{\varepsilon}/\sqrt{2\alpha}$.

Note that,
we shall only consider the case where $\sigma_{\delta x}$ is small
since $\alpha=cste$ requires essentially $\delta x$ to be small.\\
%{\bf if the shape of $J$ is not too complicated : Inutile, c'est un
%developpement de Taylor au 1er ordre, peu importe la forme de J tant
%que $\delta x$ est suffisamment petit}.\\

%In other words, let us consider a simple curve $J$, that starts from
%$J(x=0)=0$, grows linearly and reaches a maximum at $J_{m}$
%for $x=x_{m}$. One gets approximately $J^{(1)}\approx J_{m}/x_{m}$
%so $\tau_c\approx x_{m}/J_{m}$,
%$\alpha\approx 2J^{(1)}\approx 2J_{m}/x_{m}$.

So, one obtains:
\begin{equation}
\sigma_{\delta x}^2=\sigma_{\varepsilon}^2 t,
\end{equation}
if $\tau_c\ll t\ll\alpha^{-1}$. The width of the
distribution of $\delta x$ then spreads like $\sqrt{t}$ as in a
diffusion process until $t\approx\alpha^{-1}$.
As $t\gg \alpha^{-1}$, the width of the distribution of $\delta x$
around a stable position is
\begin{equation}
  \sigma_{\delta x}=\frac{\sigma_\varepsilon}{\sqrt{2\alpha}}.
  \label{ecarttype}
\end{equation}

%\begin{figure}[!h]
%  \begin{center}
%    \begin{minipage}{65mm}
%      \psfig{file=ajustement_lineaire.ps,width=65mm,angle=-90}
%    \end{minipage}
%    \caption{\small Zoom of Figure \ref{critbif} around the maximum of
%    the flux function. The flux function is fitted with three straight
%    lines with different slopes.}
%    \label{approx}
%  \end{center}
%\end{figure}

\subsubsection{In the vicinity of the bifurcation}

Let us assume that the stable state of equilibrium is in the vicinity
of $x_m$.
If $x_l$ and $x_r$ are close to $x_m$ such that $J$ is almost constant,
then $J(x_l)\approx J(x_r)$ and $\alpha\approx cste\ll 1$, one obtains
(when $t\ll\alpha^{-1}$):
\begin{equation}
  \sigma_{\delta x}^2\approx \sigma_{\varepsilon}^2 t.
\end{equation}
$\delta x$ undergoes a diffusion process and
nothing prevents from the spreading of the perturbation as long as
$\delta x$ is small enough, that is, as long as $\sigma_{\delta x}$ is
smaller than the size of the region where $\alpha\ll 1$.
In this case, the diffusion coefficient writes
$D=\sigma_{\varepsilon}^2/2$. The time needed by the perturbation to
reach any small distance $d$ from an initial position is:
\begin{equation}
  t_d\approx\frac{d^2}{D}=\frac{2d^2}{\sigma_{\varepsilon}^2}.
\end{equation}
So, if two stable states are close enough, the sytem will pass from 
one stable state to the other with a frequency equal to
$\sigma_\varepsilon^2/2d^2$. diffusion has then two effetct: the
broadening of each steady state and the jump from one steady state to
the other.\\

In the case of a supercritical bifurcation, we may try to
define when the two "steady" states remain separated and detectable
as such. It occurs when the standard-deviation (due to noise) of a
single steady state (without noise) is smaller than the distance
between the two steady states $x_l=x_r=x_{tot}/2$ and $x_m$. As the
standard deviation $\sigma_{\delta x}$ is given by equation
\ref{ecarttype}, it imposes:
%and if
%$\sigma_{\delta x}=\frac{\sigma_\varepsilon}{\sqrt{2\alpha}}$ when
%both $v<0$ and $v>0$, the experimental noise might make the
%detection of the bifurcation a hard task if the stable states
%before and after the bifurcation are not too
%distant. However, one can write a condition on the shape of the flux
%function when $v<0$ to detect the 
%bifurcation: the standart deviation $\sigma_{\delta x}$ has to be
%smaller than the distance between the stable state before the
%bifurcation $x_l=x_r=x_{tot}/2$ and $x_m$:
\begin{equation}
  \sigma_{\delta x}^2=\frac{\sigma_\varepsilon^2}{2\alpha}<(x_m-x_l)^2.
\end{equation}
With $x_m-x_l=-v/2$ and
$\alpha=2J^{(1)}(x_{l/r})\approx \frac{1}{2}vJ^{(2)}(x_m)$, one obtains:
\begin{eqnarray}
  \frac{\sigma_\varepsilon^2}{2vJ^{(2)}(x_m)}&<&\frac{v^2}{4},\\
  \Rightarrow \sigma_\varepsilon^2&<&\frac{1}{2}v^3J^{(2)}(x_m).
\end{eqnarray}
If this inequality is not satisfied, the two branches of stable
solutions in figure \ref{critbif} can not be observed and the
bifurcation is not detected as long as $v$ is too small.\\

%In the case of a supercritical bifurcation and $v=x_l+x_r-2x_m>0$, the
%distance $d$ between a stable branch and the frontier of its basin of
%attraction is $d=\sqrt{\mu}$.
%The two branches of stable solutions in Figure \ref{critbif} should be
%well observed if the time of observation of these solutions is large
%enough that is, everywhere but in a close neighbourhood of
%$v=0$. While $v$ is small it is difficult to
%distinguish between the two solutions because of the noise and thus to
%identify the type of the bifurcation {\color{blue}as long as
%$\alpha\ll 1$}.\\

%{\color{red}Attention il faut donner la largeur de d ou de mu en
%fonction de ?, l\`a o\`u on doit trouver la solution, gr\^ace \`a la pente.}\\
%{\color{blue}En fait, lorsque $v=0$,la largeur de la zone de diffusion
%  depend de la somme des pentes autour de $x_m$ : $\alpha$. Il faut
%  que $\alpha\ll 1$ pour que le syst\`eme diffuse. Au-del\`a, le
%  syst\`eme est ramen\'e pr\`es de $x_m$.}\\

In the same way, when a subcritical bifurcation occurs, as displayed
in figure \ref{sscritbif}, the jump from the single stable solution
$u=0$ to one of the two others may occur before the maximum of the
flux function, i.e. before $v=0$, so that for strong enough noise or
little subcriticality, the bifurcation may look as a supercritical
one.
%two
%unstable branches ($v\in [v_{min};0]$) may seem to merge with the stable
%branch ($u=0$ and $v<0$) if the time of observation of this latter
%stable solution is small enough. In this case, the subcriticality
%disappears because of the noise and it only remains the
%supercritical behaviour of the bifurcation.\\

Finally, if one considers a hypercritical bifurcation, the flux
function is symmetric with respect to its maximum,
$\alpha$ is always null and $J(x_l)=J(x_r)$. All the states are
equilibrium ones and, because of the noise, $x_l$ and $x_r$ constantly
fluctuates between all of them.
As a conclusion the nature the noise can hide the true nature of the
bifurcation.

\subsubsection{Far from the bifurcation}

On the other hand, if the stable state of equilibrium corresponds to
different $x_l$ and $x_r$ such that
%$x_l$ is on the left wing of Figure \ref{approx}
%while $x_r$ is on the right and that
$\alpha\approx cste\gg 1$, the
variance $\sigma_{\delta x}^2$ tends to a finite value:
\begin{equation}
  \sigma_{\delta x}^2\approx\sigma_{\varepsilon}^2/2\alpha.
\end{equation}
It means that the spreading of $\delta x$ due to the noise is balanced
by the damping due to $\alpha$. Assuming that $\alpha$ is constant,
$\delta x$ is a Gaussian random variable having the following
probability density function:
\begin{eqnarray}
  f_{\delta x}(X)&=&\frac{1}{\sigma_{\delta x}\sqrt{2\pi}}e^{-\frac{X^2}{2\sigma_{\delta x}^2}},\\
  &=&\sqrt{\frac{\alpha}{\pi\sigma_\varepsilon^2}}e^{-\frac{\alpha
  X^2}{\sigma_\varepsilon^2}}. \label{proba}
\end{eqnarray}
The probability to jump from a stable solution to another one is the
probability to have $\delta x$ larger than the distance $d$ from the
initial stable state to the frontier of its basin of attraction. As
$\alpha$ has a damping effect, one expects that the probability for a
jump to occur should decrease as $\alpha$ increases. From equation
\eqref{proba}, one can write this probability:
\begin{equation}
  P(\delta x>d)=\frac{1}{2}\left [ 1-\mbox{erf}\left
  (\frac{\sqrt{2}\alpha d}{\sigma_\varepsilon}\right )\right ],
\end{equation}
where $\mbox{erf}(.)$ denotes the error function and $P(\delta x>d)$
decreases as $\alpha$ increases as expected.

In fact, the latter result can be applied if a supercritical or a
subcritical bifurcation has occured and not after a hypercritical
bifurcation.
%In this case, if $x_l$ and $x_r$ are very different and
%$J(x_{l/r})$ changes quickly,
After a hypercritical bifurcation the only result one can write is
that the life expectancy, $t_l$, of a stable state is proportional to
the time needed by the noise to make the two boxes exchange one
particle:
\begin{eqnarray}
  t_l&\propto&\frac{1}{\sigma_{\varepsilon}^2},\\
  &\propto&\frac{1}{J(x_l)+J(x_r)}.
\end{eqnarray}
So, as $J(x_l)+J(x_r)$ tends to zero, the corresponding equilibrium
state will be observed during a time much longer than the one for
which  $J(x_l)+J(x_r)$ is large. Thus, from an experimental point of
view, much attention has to be paid
to the interpretation of the experimental results. Indeed, a state may
be wrongly interpreted as a non equilibrium one if
$J(x_l)+J(x_r)$ is large.

\section{Experimental part}\label{exp}

A well-known experiment involving a system of two communicating
containers partially filled with macroscopic spherical particles is
the so-called granular ``Maxwell demon''.
When such compartments are shaken together at same frequency and
amplitude, it has been often quoted that, under some
circumstances, one compartment empties spontaneously in the other
\cite{schlichting96}.
This symmetry breaking comes from the fact that the outgoing flux
function is not monotonic as the number of particles increases: it
first grows as the number of beads $N$ increases and, as
$N>N_{max}$, decreases because the dissipation becomes more
important. Then, it seems to be a relevant experiment to test the
prediction of the previous section. Note
that this experiment has been studied in \cite{eggers99}. He
used the equality of the outgoing flux from each box as the
equilibrium condition, and
a continuous model taking into account that the effective
granular temperature decreases when the density of particles
increases. He concludes that the symmetry breaking follows a second
order phase transition. However, it has been pointed out that the
Egger's approach suffers some discrepancies with respect to
experiments \cite{pierre02} and that a thermodynamics analogy to
describe such a system is questionable \cite{pierre02bis}.
An other approach has been proposed
\cite{brey01} when the size of the hole connecting both compartments
is larger than the mean free path of the particles. It is based on the
pressure balance between the two boxes and leads to the conclusion
that beyond a critical value of the number of particles, the symmetry
breaks spontaneously. Anyway, this surprising behaviour has motivated
many experiments using mixtures of different particles, different
boundaries conditions or different number of compartments for instance
\cite{lipowski02,barrat03_2,mikkelsen05,vandermeer06}.
But the influence of the experimental parameters on the shape of the
flux function, and then on the kind of bifurcation leading to the
symmetry breaking, has not yet been addressed.\\

In the present experiment, we study a granular Maxwell demon
experiment using the outgoing flux function $J$ to forecast what kind
of transition occurs when the symmetry breaks. The control parameter is
the number of particles $N$ and the analysis is made for three different
frequencies. We use a single vibrated box to measure $J$ as a function
of the number of particles $N$ that it contains and of the frequency of
excitation. Our analysis is based on the same equilibrium condition
than the one used in \cite{eggers99}, that is on the balance of
the outgoing flux between the
two compartments. Note that because the number of balls is discrete,
there is an intrinsic noise in our experiments but we will see that it
is nevertheless possible to identify different bifurcations.
It is also important to note that in the case of two
vibrating boxes and if $N$ is large enough, interactions between particles
coming from one compartment to the other could occur in the
vicinity of the route between the boxes. However, this latter would
occur at much larger densities. It would lead in this case to a flux
function depending on the number of particles in both boxes which is not
our purpose here.

\subsection{Experimental apparatus}

\begin{figure}[!h]
    \begin{minipage}{75mm}
      \begin{center}
      \psfig{file=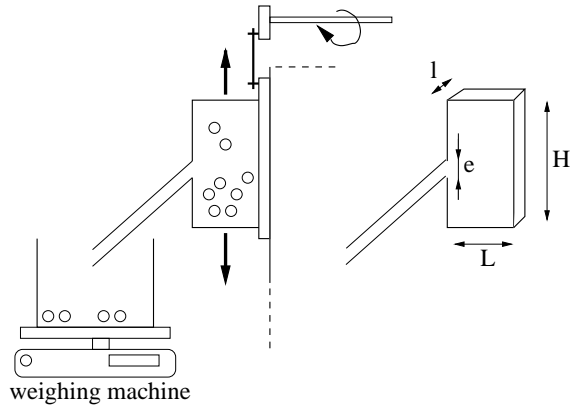,width=75mm,angle=0}
      \end{center}
    \end{minipage}
    \caption{\small Experimental apparatus.}
    \label{dispo}
\end{figure}

The experimental apparatus is shown on Figure \ref{dispo}.
A single box of dimension $L=20mm$, $H=30mm$, $l=12mm$, with a narrow
slit on one side at height $10mm$ and of width $3mm$, is partially
filled with $N\approx 680$ steel spheres of diameter $1mm$.

The box is fixed on a sliding girder connected to a crankshaft and a
rotating wheel. It allows to impose a linear oscillation to the box,
$A\sin{(2\pi f t)}$ where $f$ and $A$ are the frequency and the
amplitude of the sinusoidal oscillation respectively:
$f\in [26Hz-43Hz]$ and $A=1.25\ mm$. The mechanical device is mounted
on a massive steel socle in order to avoid any perturbation.
\begin{figure}[!h]
  \begin{center}
    \begin{minipage}{80mm}
      \psfig{file=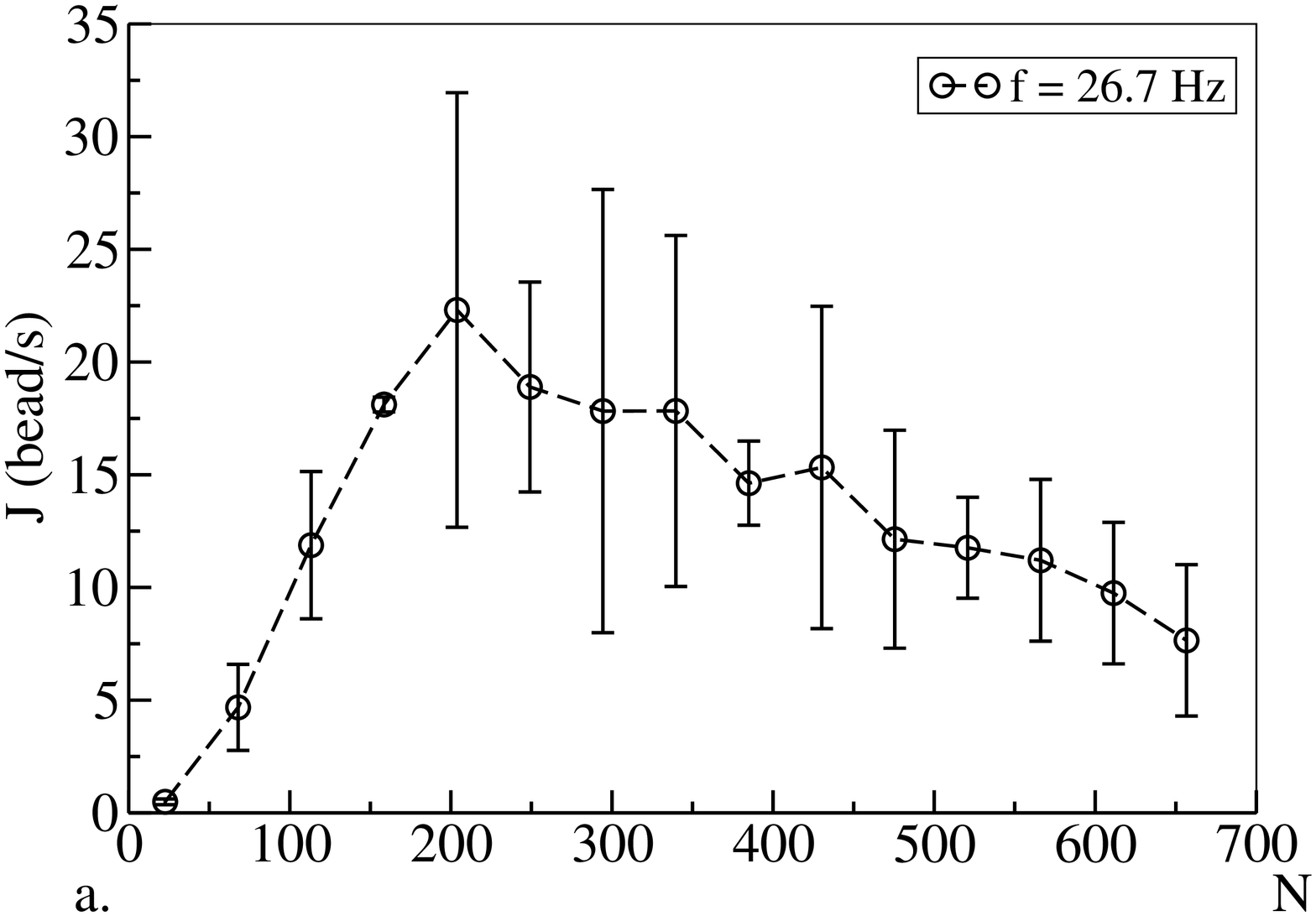,width=65mm,angle=-90}
    \end{minipage}
    \begin{minipage}{80mm}
      \psfig{file=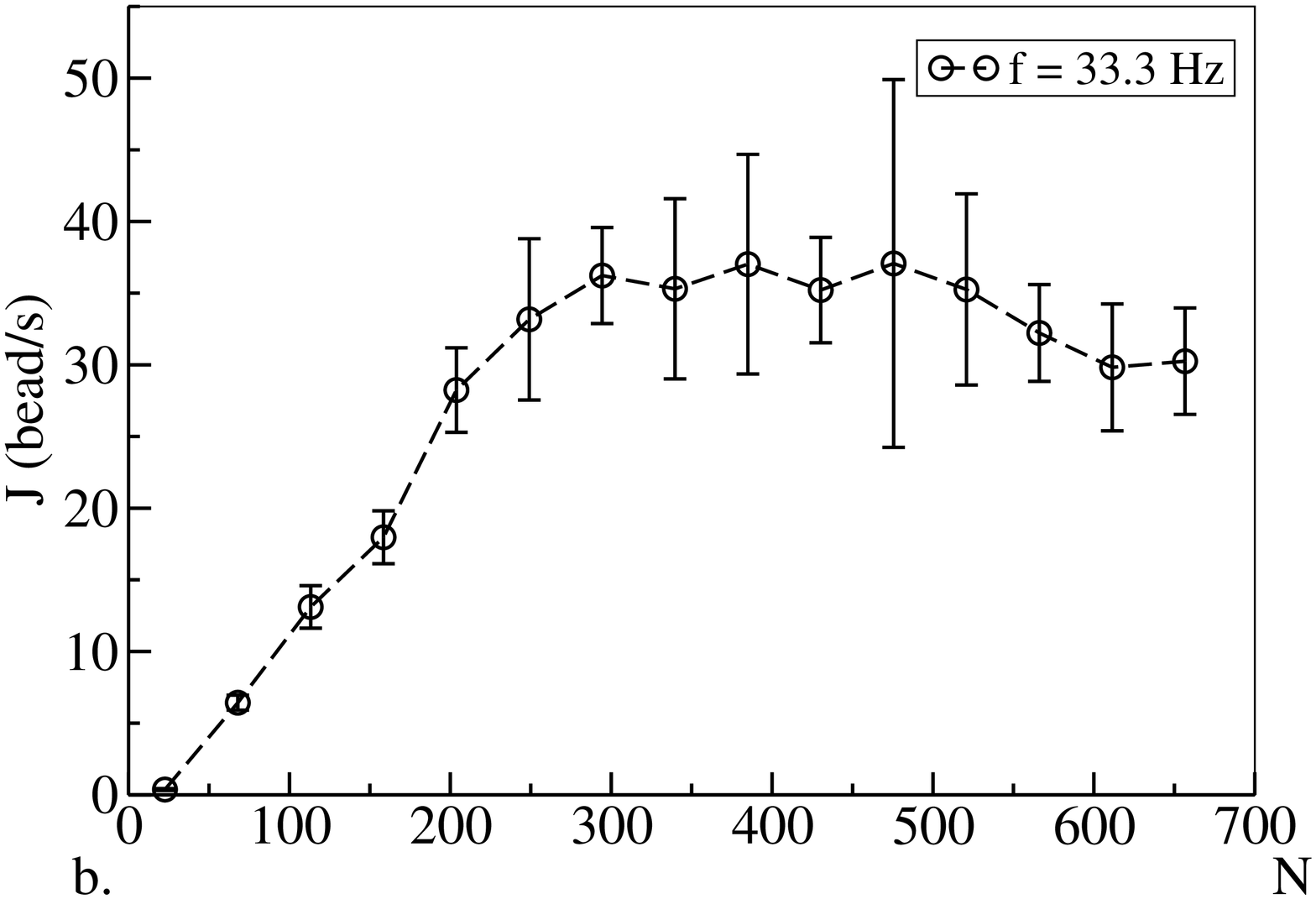,width=65mm,angle=-90}
    \end{minipage}
    \begin{minipage}{80mm}
      \psfig{file=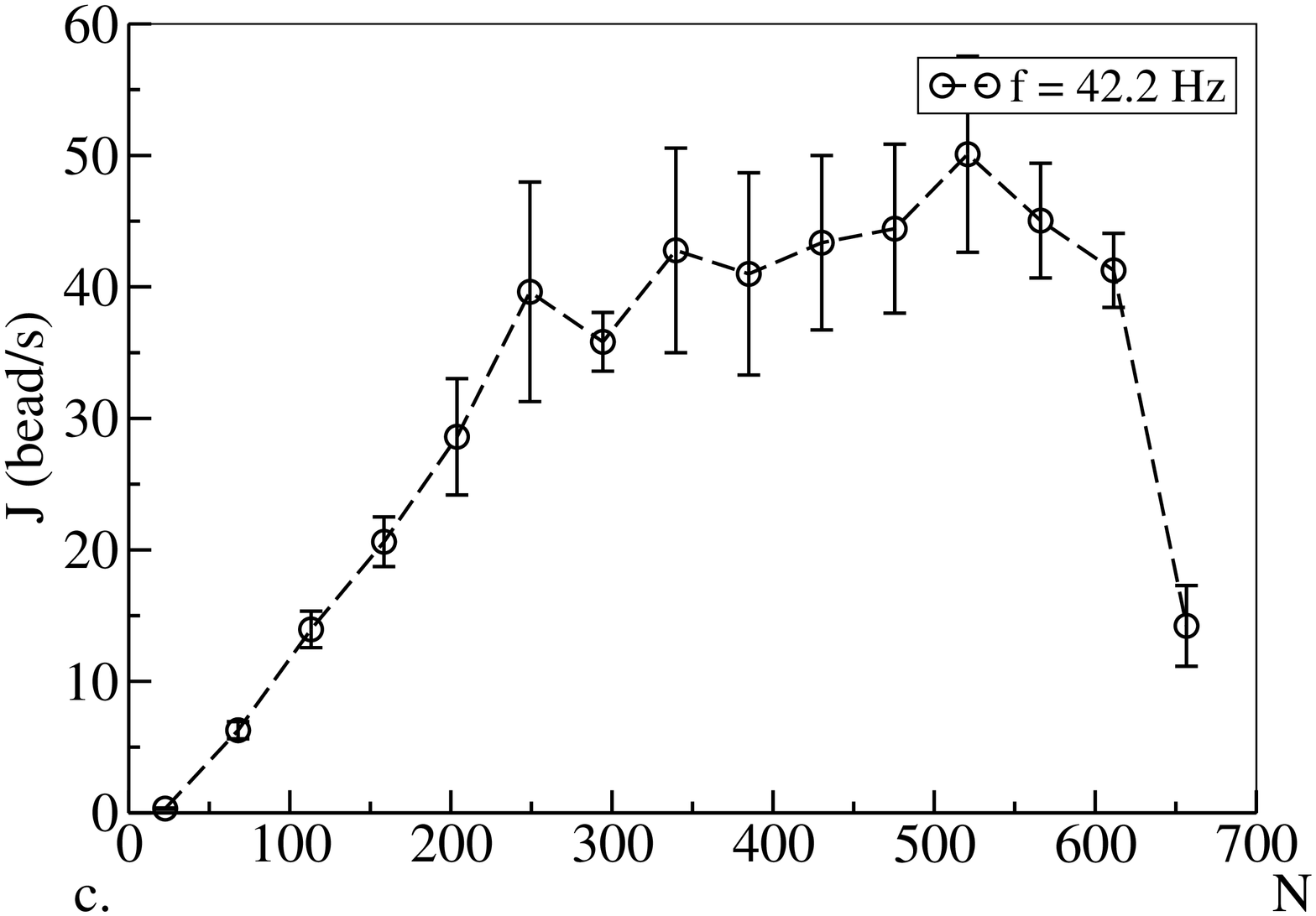,width=65mm,angle=-90}
    \end{minipage}
    \caption{\small Experimental flux as a function of the number of
      particles in the container shaked at frequency
      $f$. a. $f=26.7Hz$, b. $f=33.3 Hz$, c. $f= 42.2 Hz$.}
    \label{courbes_flux_exp}
  \end{center}
\end{figure}

The frequency is imposed with an electrical motor and controlled, during
each experiment, thanks to a tachometer to measure the fluctuaction of
the frequency. The maximum amplitude of these fluctuations is
smaller than $2\%$ of the imposed frequency. When
beads leave the vibrating compartment, a long tube bring them to a
repository with a soft bottom such that grains do not bounce. The
evolution of the mass of the content's repository is measured each
$0.5s$ thanks to a weighing machine with a precision much less than
$10^{-3}g$ (which is the tool accuracy), because of parasite effect
that includes mechanical vibration and ball collision with
bottom. Data are then recorded with LabVIEW Software on a
computer and smoothened. For a given frequency $f$, the outgoing flux $J(t)$ is
derived from the temporal mass evolution. This allows to finally plot
$J$ as a function of the number of beads, $N$, remaining in the vibrating box.

\subsection{Experimental results}

Figure \ref{courbes_flux_exp} shows the outgoing flux, $J(N)$,
obtained for three imposed frequencies $f=26.7 Hz$, $33.3 Hz$ and
$42.2 Hz$. Each plotted curve is the mean of several
experiments. They all starts from $J(0)=0$, increases until it
reaches a maximum and then starts to decrease. This behaviour reminds
us of the different shapes of flux function we studied in the
previous section.

One can already note that Figure \ref{courbes_flux_exp}.a
has a left wing steeper than the right wing and that
\ref{courbes_flux_exp}.b displays a right wing steeper than the left
one. So, if the experiments are described correctly by the model of
the previous section and even with some noise, we may be tempted to
conclude that the kind of bifurcation changes as the frequency of
vibration increases.

Assuming a Gaussian white noise, the standard deviation (s.d.) is
computed by multiplying the empirical s.d. with the $5\%$-quantile of
the Student law with $p$ degrees of liberty, where $p$ is the number
of experiments \cite{bouquin_stat}. Indeed, it is a classical way to
proceed when the number of experiments is small. As one can expect, at
fixed frequency, the uncertainty is large when the flux is large. Note
that three experiments have been made for $f=26.7Hz$ and nine
experiments have been achieved for $f=33.3Hz$ and $f=42.2Hz$. This
explains why the uncertainties seem larger for $f=26.7Hz$ than for the
two other frequencies used.\\

One has also to mention that in these experiments and for
the largest values of $N$, the
number $n$ of layers of particles is $n\approx 3$. The distance
between the top layer and the slit is then $D\approx 7\ mm$. For the
highest frequency, $f=42.2\ Hz$, the height reached by a particle
initially at rest is $h\approx 5.6\ mm$. As $h\approx D$, the density
of particles in the vicinity of the slit might be sufficiently large
such that interactions occur between the beads of the two boxes. It means
that the tail of each curve in Figure \ref{courbes_flux_exp}, that is
for the largest number of particles, could
have been modified by the high concentration of particles near the
slit. In this case, this would involve another equilibrium condition
than the one on flux and then an approach using fluid parameter,
similar to \cite{brey01}
for instance, could be more suitable to this problem.\\

Anyway, if one assumes that interactions do not occur, one can use the
result of section \ref{bifurtheo}. Each curve is then fitted to a
polynomial function of degree chosen thanks to a likelihood ratio test
\cite{bouquin_stat}. One then computes the value $N_m$ for which
$dJ/dN=0$ and the successive derivatives of $J$ at $N_m$. The
result is that Figures \ref{courbes_flux_exp}.a,
\ref{courbes_flux_exp}.b and \ref{courbes_flux_exp}.c correspond to
supercritical, hypercritical and subcritical bifurcations
respectively as $N$ increases. As forecast previously,
the kind of bifurcation depends on the frequency used during an
experiment.\\
Let us discuss briefly how the system converge to an
equilibrium stable state and the effect of the noise. The convergence
to a stable state depends on
the sum of the derivatives of the flux function at $N_l$ and
$N_r$. However, as  $N$ is a discrete value, the flux of $N$ is
discrete and fluctuates, and the exact equilibrium state is not
reachable. It results a standard deviation at $N_{l}$ and $N_r$ which
is inversely proportional to the sum of both slopes around $N_l$ and
$N_r$ (see equation \ref{ecarttype}).
%the exact equilibrium
%state is not reachable {\color{blue}introducing an {\it intrisic}
%noise. Indeed, this {\it intrinsic}
%noise due to the discrete nature of the system induces a noise on the
%flux function proportional to the sum of both slopes around $N_l$ and
%$N_r$ leading to a feedback on the fluctuation around
%$N_{l/r}$. Consequently, one concludes that the larger the sum of the
%slopes of the flux function at $N_{l}$ and $N_r$, the larger the
%fluctuation around $N_{l/r}$ due to the discreteness of the system.}
%In this case, the system fluctuates around
%the equilibrium state with an intensity given by the sum of both
%slopes around $N_l$ and $N_r$. So, if this sum is large, the fluctuations
%of $N$ are large. {\color{red}(Citer Equation)}
%Inversely, if the flux does not vary much
%around $N_l$ and $N_r$, the fluctuations around a stable state are
%small and can even reduce to one particle. Obviously, the experimental
%noise adds to the fluctuations described above.\\

%Note also that if $N_l=N_r$ is a stable state, the convergence
%depends on $\alpha=J^{(1)}(N_l)+J^{(1)}(N_r)$ but the critical slowing down
%near this state is limited because $N$ is a discrete quantity.
%In the particular case where $N_l=N_r$ is a stable state of
%equilibrium, the time of convergence to this state decreases as $N$
%tends to $N_l$ because $J^{(1)}(N_l)+J^{(1)}(N_r)$ tends to $0$. If
%the equilibrium state is reached for $N_l\neq N_r$,
%decreasing as $J$ tends to its maximum.\\

Recently,  Mikkelsen {\it et al} \cite{mikkelsen05} studied a granular
Maxwell demon experiment and the effect of statistical fluctuation
on critical phenomena.
Their numerical simulations show qualitatively the same kind of flux
curves as ours, as the frequency increases.
The symmetry breaking is studied with the frequency as the control
parameter and with different total number of particles $N$ to highlight
the effect of a noise. They find that, at fixed $N$, the
transition is supercritical when the frequency becomes larger than a
critical threshold. However, according to our analysis, it can be
either supercritical, subcritical or hypercritical when $N$ becomes
larger than a threshold, the type of bifurcation depending on the
frequency of vibration.
%As one can expect, it shows that the kind of bifurcation
%might depend on the control parameter studied.
They also find that they are faced with a supercritical bifurcation when the
flux curve is symmetric whereas
our study proves that the bifurcation is hypercritical and that the
system is driven by the noise in this case. This shows how the type of
bifurcation depends on the control parameter used.\\
As a final remark, it is worth mentionning that a Maxwell demon 
experiment checks directly the breaking of symmetry. It tests then the
shape of the flux function near its maximum and cannot then be
considered as a proof of any modelling of the flux function far from
this maximum. Consequently, most of the flux function assumed in the
litterature cannot be considered as settled on experimental evidence.

%and also that it is
%particularly tricky to conclude on the type of an
%experimentally observed bifurcation when the noise is large.

\section{Conclusion}\label{conclu}

In conclusion, we have proposed a theoretical analysis of the behaviour
of two subsystems able to exchange each other some part of their
internal quantity thanks to a flux function $J$. We have shown that when
$J$ is not monotonic, a bifurcation can occur leading to different
equilibrium states. The type of bifurcation is then found to depend on
the detail of the shape of the flux function near its maximum. In
particular, we have pointed out a new kind of bifurcation called
hypercritical when the shape of the flux function is
symmetric with
respect to its maximum. In this case, a continuum of solution exists
making the dynamics of the system govern by the noise. In the other
cases, the noise, if large enough, is found to have an impact on the
conclusion concerning the kind of bifurcation which appears.
% all states are states of equilibrium. The effect of the noise has been
% addressed and has been proved to have an impact on the conclusion
% concerning the kind of transition which appears.
In the last part of this paper, we have presented a simple granular
Maxwell demon experiment to
investigate the kind of transition occurring when the symmetry
breaks. Unlike previous experiments, we use the number of
particles as the control parameter with different frequencies of
vibration $f$. Our
results show that different bifurcation, including the hypercritical
one, occur depending on $f$. This behaviour was not pointed out before
neither experimentally, nor numerically, not theoretically.
It
%We have concluded that it depends on the parameters of excitation and
%more specifically on the frequency of vibration. This behaviour
could then allow to investigate new kind of bifurcation.\\
However the detection of the true nature of bifurcation
(supercritical vs. subcritical, etc) is made difficult due to the
little number of grains (and by the noise it induces) in the case of
a Maxwell demon experiment. This is easier if one uses directly the
flux function which gives directly the symmetry of the curve. In
turn this means the use of a single half box. We believe that a
Maxwell demon experiment with $100$ balls only test the presence of a
bifurcation, but not the precise shape of the flux function near its
maximum or further in the tails. So,
previous Maxwell demon experiments can not be considered as a
test of the validity of any modelling of the flux function far from
this maximum as it is sometimes done in the literature.\\

%As a final remark, it is worth mentionning that a Maxwell demon
%experiment tests directly the breaking of symmetry through the shape
%of the flux function near its maximum. It can not then be considered
%as a test of the validity of any modelling of the flux function far
%from this maximum as it is sometimes done in the literature.}\\

%{\color{red}fait : partie I et II, III, IV et V

%1. Nature ? Maxwell demon in GG as a paradigm for new bifurcation kinds ?
%voir pour les fluctuations experimentales et le bruit : 
%bruit proportionnel à la pente ?  OK\\
%2. Mais =>donner l'extension de la zone de diffusion pour v=0, et
%sigma donn\'e.}{\color{blue} Fait. La zone de diffusion d\'epend des pentes
%autour de $x_m$, elle s'\'etend sur une zone telle que $\alpha \ll 1$
%et est donc propre \`a chaque courbe de flux.}\\
%{\color{red}3. On ne parle pas de ralentissement
%  critique.}{\color{blue} Le ralentissement
%critique est discut\'e \`a la suite de l'\'equation 12.}\\
%{\color{red}4. D\'ecrire le cas v>0 et hypercritique : il n'y a pas
%de solution, car
%le flux est <0 pour N correspondant à v>0 (puisque J=0 à N=0 => le
%mod\`ele n'est pas r\'ealiste => bonne intro pour le §
%suivant).}{\color{blue} OK, phrases ajout\'ees \`a la fin du
%  paragraphe sur le cas sym\'etrique.}

{\bf Acknowledgements:}\\

Centre National d'Etudes Spatiales (CNES) is gratefully thanked for
financial support.

\end{document}